# Mass transport driving forces under electric current in the liquid Sn-Zn system

Jean-Philippe Monchoux[1,*]

[1] CEMES, Université de Toulouse, CNRS, 29 rue Jeanne Marvig, BP 94347, 31055 Toulouse, France

[*] Corresponding author: monchoux@cemes.fr

**Abstract**

Significant effects of electric currents on mass transport in liquid metals have been observed for long, but the origin of the corresponding driving forces remains unclear in the literature. Without current, two driving forces induce mass transport in liquid metals. (i) A chemical force, coming from concentration gradients. In that case, mass transport occurs by diffusion. (ii) A physical force, resulting from density gradients thermally and/or chemically induced. Here, mass transport occurs by thermal and/or solutal convection. Under electric currents, these driving forces are modified, either by electrostatic or magnetic forces, the corresponding mechanisms being referred to as electroconvection and magnetoconvection, respectively. However, these mechanisms cannot easily be distinguished from each other, leading to confusion in literature. Here, it has been shown that, in the liquid Sn-Zn system, the driving force induced by 500-1000 A/cm$^2$ electric current densities is magnetic rather than electrostatic, the mechanism being therefore magnetoconvection.

The influence of electric currents on metallurgical phenomena have been the subject of numerous studies since the discovery of electromigration in 1861 by Gerardin [1, 2]. The effect of $\approx 10^2$ A/cm$^2$ electric currents on mass transport is particularly striking in liquid metals, where increase by factors of 10 and above in diffusion coefficients have been observed [3]. To account for this phenomena, different approaches were proposed. Because the mass transport phenomena in liquid may have different origins, on which the current can add different contributions, it is probably necessary to make here a short summary of the various induced phenomena, which will later be analyzed in more details. Consider first the mass transport mechanisms in liquids without the action of an electric current. In one hand, when mass transport occurs under chemical forces coming from concentration gradients, diffusion following Fick's law is induced. But because the system is liquid, mass transport can also occur under the action of physical forces resulting from density gradients in the liquid, either thermally or chemically induced, giving rise to thermal and solutal convection, respectively. With the action of an electric current, additional contributions can be superimposed to these driving forces. First, electrostatic forces can be induced by the electric field (electron wind effect). If these forces modify the diffusion processes without inducing convection, they are classically referred to as electromigration forces. On the contrary, if they activate convection, they are called electroconvection forces. But magnetic forces can also be generated due to motion of the liquid in a magnetic field, giving rise to Lorentz forces. In that case, the corresponding effect can be called magnetoconvection.

Pioneering but also recent works consider only electrostatic forces, generated by current densities of the order of ≈ $10^2$-$10^3$ A/cm$^2$ [3-6]. Among them, some studies consider that these forces generate electromigration [5], other propose the activation of electroconvection [3, 4]. Moreover, because poorly understood convection processes were involved, the action of "unknown" mechanisms has also been proposed [7]. However, recent works on liquid metal batteries have accurately described magnetoconvection processes under the action of the magnetic force, in the case of current densities of the order 100 mA/cm$^2$ [8-10]. In these studies, the possible action of the electrostatic force is not discussed, even in the extensive review of Kelley et Weier [9]. Moreover, under current densities of the order of ≈ $10^2$-$10^3$ A/cm$^2$, the driving force controlling mass transport in liquid metals remains unclear. Notably, its order of magnitude is unknown.

Therefore, a method has been employed here to decouple the different contributions of mass transport in liquid metals submitted to ≈ $10^2$-$10^3$ A/cm$^2$ electric currents, in order to determine the corresponding driving forces. In the absence of electric current, in the case of a solid metallic solute dissolving in a liquid metal, the flux of solutes in a concentration gradient following the classical Fick's law gives a concentration profile following the expression [11]:

$$C(x) = C_\infty + (C_i - C_\infty) \cdot \left[1 - \mathrm{Erf}\left(x/2 \cdot \sqrt{D_L(t+t')}\right)\right] \qquad [1]$$

with $C(x)$: concentration at distance $x$ of the source of solutes, $C_i$: concentration at the interface between the source of solutes and the liquid, $C_\infty$: concentration at infinity, $D_L$: solute diffusion coefficient in the liquid, $t$: duration of the isothermal plateau, and $t'$: time during which Sn is in the liquid state in the heating and cooling steps (≈ 1 min in total).

Still without current, thermally and/or chemically induced convection activates stirring of the liquid, and modifies strongly the diffusion profiles given by Equ. [1]. Examples of theoretical solute concentration maps under the action of convective processes can be found in the work of Personnettaz *et al.* [12], showing that the concentration of the solute does not follow the classical Fick's law. Convection can first be activated by thermal gradients, which create density gradients in the liquid. The results presented in our study show that the experimental setup used effectively eliminated this contribution, which will therefore not be taken into account. But convection can also be caused by concentration gradients coming from the dissolution of a solid into a liquid (solutal convection), which generates Archimedes' forces per unit volume $\vec{F_A}$ [13]:

$$\vec{F_A} = (\rho_i - \rho_\infty)\vec{g} \qquad [2]$$

with $\rho_i$: density of the liquid at the solid-liquid interface, $\rho_\infty$: density of the liquid at infinity, and $\vec{g}$: gravity vector.

In the presence of an electric current, two driving forces are generated, as stated above. First, the force induced by the electron wind, resulting from ballistic momentum transfer from the electrons to the moving ions in the liquid, which is added to the force created by an electric field on charged ions. Therefore, an effective electrostatic force per metallic ion $\vec{f_e}$ can be derived, as follows [11]:

$$\vec{f_e} = Z^*|e|\vec{E} \qquad [3]$$

with $|e|$: electron charge, $\vec{E}$: electric field created by the electrical resistance of the liquid following the Ohm's law, and $Z^*$: effective charge, which takes into account the ballistic and (purely) electrostatic contributions. This force generates either electromigration (if mass transport rate is modified without activation of convection) or electroconvection (if mass transport is modified by activation of convection).

The second driving force is the Lorentz force per metallic ion $\vec{f_L}$, which is created by the magnetic field $\vec{B}$, generated by the current flowing through the liquid capillary, on the ion of charge $q$ moving at velocity $\vec{v}$:

$$\vec{f_L} = q \cdot \vec{v} \wedge \vec{B} \qquad [4]$$

This force generates convection, and thus magnetoconvection mechanisms.

Therefore, in the following, these driving forces have been evaluated quantitatively, through mass transport experiments of Zn dissolving into liquid Sn under current densities of $\approx$ 500-1000 A/cm².

For this purpose, diffusion experiments in the liquid Sn-Zn sytem under electric current were carried out, as follows. Capillary reservoirs containing liquid Sn closed in one extremity by solid Zn (Fig. 1) were employed. This allowed Zn to dissolve either on top (Fig. 1a) or at the bottom (Fig. 1b) of the liquid Sn capillaries, to activate solutal convection, or, on the contrary, to avoid it (the Zn-rich Sn liquid being of lower density than pure liquid Sn, convection provoked by the Archimedes' force is activated when Zn is in lower position). Capillaries of 0.55 mm in radius were thus drilled in stainless steel reservoirs 8 mm in diameter and 1 cm long, over a depth of 7 ± 1 mm (the capillaries were thus closed at one end, see Fig. 1). Stainless steel was selected for its very low reactivity with both Zn and Sn at the temperature of the experiments (250°C). In the capillaries, Sn wires (99.99+%, Goodfellow) 0.5 mm in radius and 8 mm long were introduced, and Zn discs (99.999%, Goodfellow) were put in contact with the reservoirs to close the capillaries.

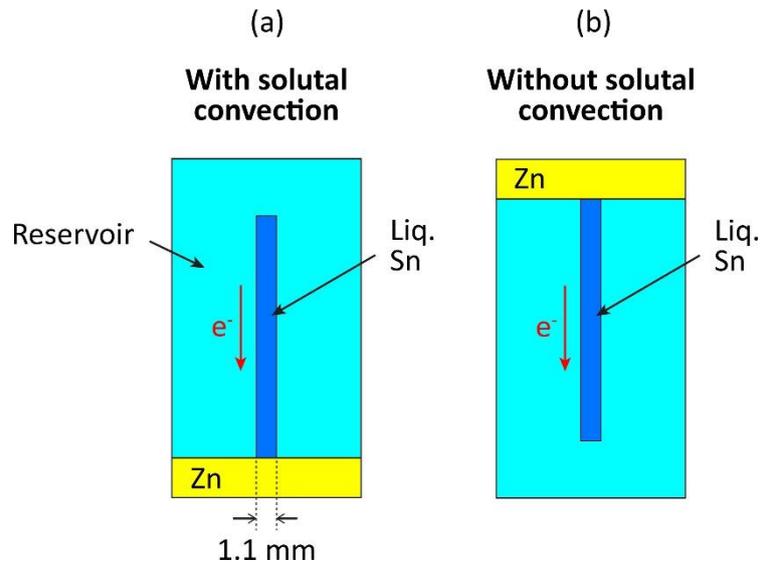

*Figure 1. Schematic drawings of the two configurations of the experiments. (a) Without solutal convection, and (b) with solutal convection.*

To apply an electric current, a spark plasma sintering (SPS) apparatus (Dr. Sinter 2080 equipment, located at PNF2-CNRS, Toulouse) has been employed, following a procedure previously developed [14]. Two configurations were employed (Fig. 2). In both configurations, graphite molds with 8.4 mm inner diameter were employed. 50 MPa uniaxial pressure was applied on the Zn discs through graphite or Cu electrodes. Consequently, the open ends of the capillaries were hermetically closed by the Zn discs, thus insuring sealing. Moreover, the Zn disks could be plastically deformed, which allowed small amounts of solid Zn to enter into the capillaries, thus accommodating slight losses of liquid Sn. In the configuration shown in Fig. 2a, the capillaries were insulated from the current delivered by the SPS by use of a PEEK polymer foil 0.2 mm thick (resisting to temperatures above 300°C) surrounding the reservoirs. However, graphite foils (papyex) 0.2 mm thick were positioned on the upper and lower ends of the graphite electrodes, to insure Joule heating at 250°C of the graphite molds by conduction of the SPS current. In the configuration of Fig. 2b, the graphite punches were replaced by Cu electrodes, and a PEEK foil was positioned around the reservoirs and Cu electrodes, to force all the current of the SPS to cross the reservoirs.

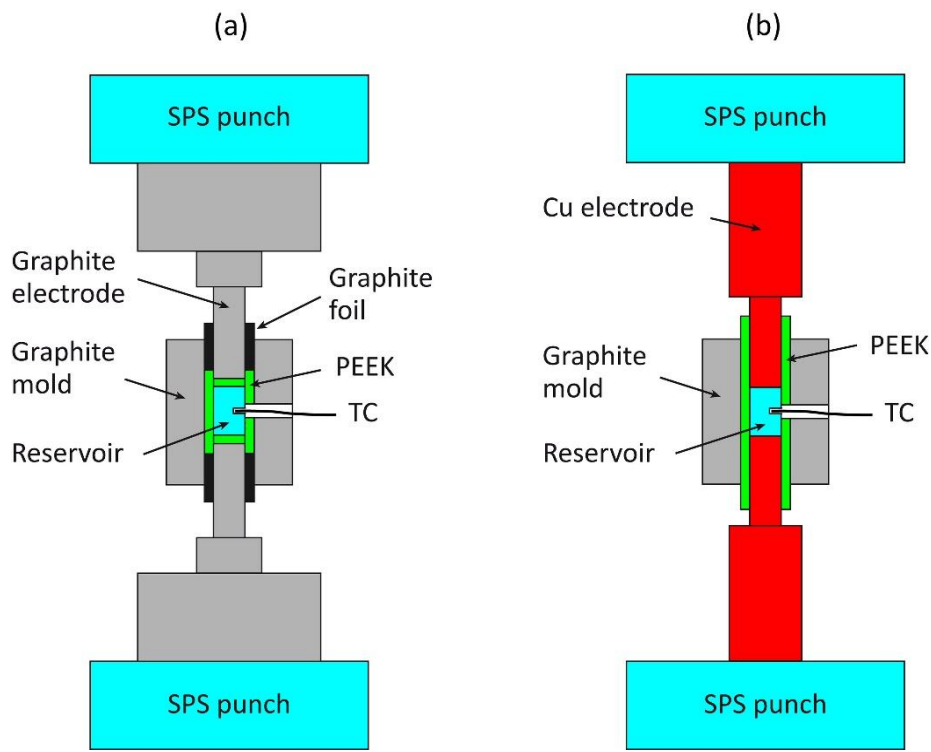

*Figure 2. Schematic drawings of the SPS experiments. (a) Experiments without current. (b) experiments with 500-1000 A/cm² currents.*

Figure 3 gives examples of current density and temperature profiles during the SPS experiments. Temperature was measured within ± 1°C by a thermocouple (TC) at 1 mm from the capillaries thanks to holes drilled in the reservoirs (see Fig. 2), and accurately monitored by a PID controller according to the following pattern: heating at 112.5°C/min for 2 minutes, heating at 25°C for 1 min, maintaining at 250°C for 1 to 10 minutes, cooling at 50°C/min for 2 minutes, and rapid quenching (in less than 30 seconds) of the graphite molds into water. This procedure was followed with care, because it has been shown that the temperature cycle, and particularly the cooling step, had significant impact on the measured diffusion kinetics in solid-state experiments [14].

The heating current delivered by the SPS was direct (no inversion of polarity), and composed of repetitions of 12 pulses of 3 ms durations separated by 6 ms of dead time. However, efficient values of this signal were recorded by the amperemeter attached to the SPS. Fig. 3 shows that the current density evolved during the constant temperature step, from about 900 A/cm$^2$ to about 700 A/cm$^2$ in the example shown. Therefore, mean current densities were calculated from integration of the profiles during the constant temperature step at 250°C, giving a value of 782 A/cm$^2$ in this example. Moreover, these average current densities exhibited values spreading between 500 A/cm$^2$ and 1000 A/cm$^2$ among the experiments. This is attributed to high sensitivity of the current density to contact resistances in the SPS setup [14]. The positive electrode of the SPS was that situated at the bottom, meaning that the electrons moved from up to down.

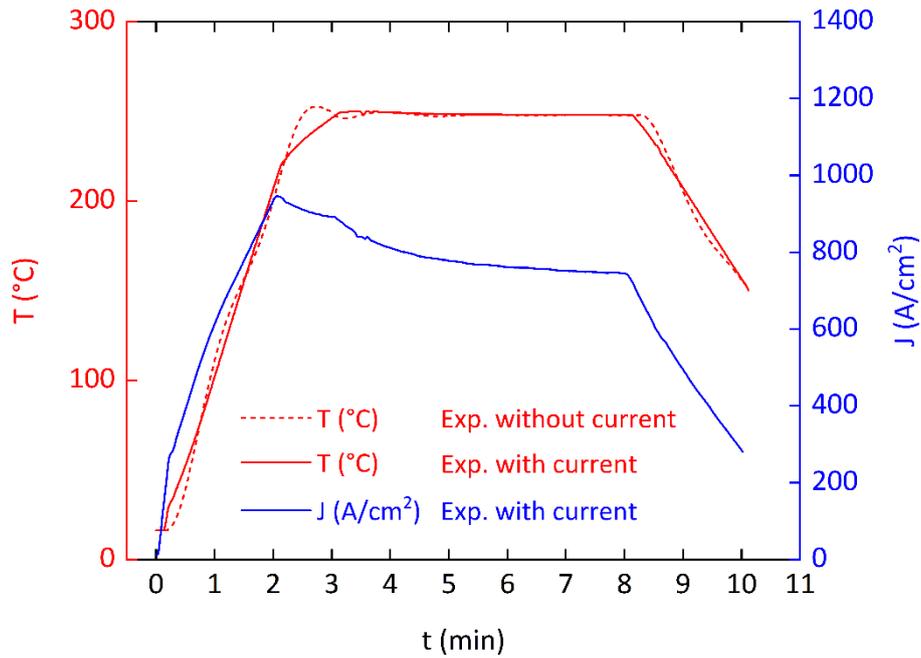

*Figure 3. Typical current and temperature profiles during an electroconvection experiment at 250°C for 5 min of holding time (Zn up configuration). Mean current density: 782 A/cm$^2$.*

After the diffusion experiments in the SPS, to analyze the concentration profiles, the capillary reservoirs were cut longitudinally, and microstructurally characterized by scanning electron microscopy (SEM) and energy dispersive X-ray analysis (EDX), using procedures detailed in the Supplementary materials. Thus, Fig. 4 shows the influence of electroconvection without contribution of solutal convection, the Zn-rich liquid floating on top of the pure Sn liquid. Without current, Zn (in yellow, Fig. 4a) diffuses over distances ranging from about 3 mm to 6 mm for 1 min. to 10 min. of diffusion time. In these cases, the concentration profiles follow the classical Erf functions given by Equ. [1], with a usual diffusion coefficient value [15] (Fig. 4c). Therefore, no convection processes, for example thermal convection, were activated in these experiments. On the contrary, when a current of 782 A/cm$^2$ is applied, the mass transport of Zn in liquid Sn is profoundly modified (Fig. 4b), leading to an almost homogeneous concentration in the capillary reservoir (Fig. 4c). Therefore, the electric current strongly activates the diffusion processes, either by electrostatic or magnetic forces.

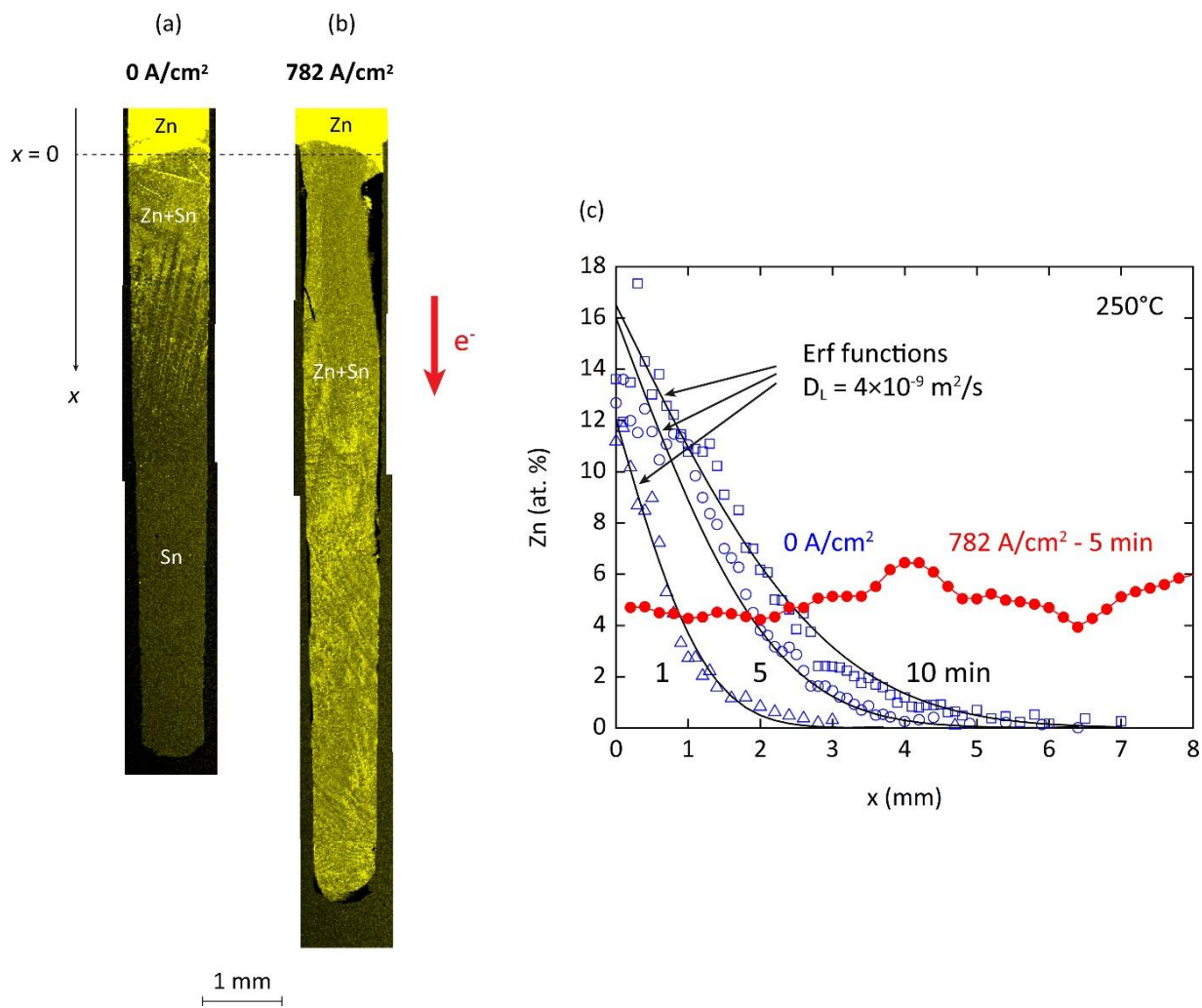

*Figure 4. SEM-EDX chemical characterizations of Zn in upper position diffusing into liquid Sn at 250°C, after solidification and cross-section metallographic preparation of capillaries. (a-b) Chemical maps of Zn (in yellow) after 5 minutes of diffusion, showing that the mass transport occurs over much longer distances with current than without current. (c) SEM-EDX linescans along the length of the capillaries (x coordinate, in mm, origin at the Zn-Sn interface), showing that the Zn concentration profiles without current (0 A/cm$^2$, in blue, 1 to 10 minutes of diffusion duration) follow classical diffusion kinetics in Erf functions (black lines), with a diffusion coefficient typical of liquids (diffusion coefficient of Zn in liquid Sn at 250°C: $D_L$ = 4-5×10$^{-9}$ m$^2$/s [15]). This shows that thermal convection is not activated in this case. On the contrary, the strong modification of the profile shape in presence of a current (782 A/cm$^2$, in red) evidences activation of convection in the liquid by the electric current.*

In the presence of solutal convection (Zn in bottom position), the concentration maps (not shown) in all cases resembled those given in Fig. 4b, meaning that the capillary reservoirs were completely filled with Zn, whether with and without current. This means that solutal convection alone (without current) leads to similar activation of mass transport than solutal convection plus contribution of the current. This effect has been more precisely quantified by measuring the total amount of Zn dissolved in the capillary reservoirs (Table 1), data from which the contributions of solutal convection and of convection under electric current have been deduced (Table 2). It can be seen that the effects of convection under electric current (whether with or without solutal convection) and solutal convection (whether with or without convection under electric current) are to increase on average the dissolution rates by factors ≈ 1.5 and ≈ 2.9 respectively. This clearly indicates that the contribution of convection under electric current to mass transport is of the same order of magnitude as solutal convection, or slightly lower in the present case.

Table 1. Amounts of dissolved Zn in liquid Sn determined by EDX (see Supplementary materials), during mass transport experiments without and with electric current (conditions for all the experiments: 250°C, 5 min).

| Configuration | Zn up 0 A/cm$^2$ | Zn up 782 A/cm$^2$ | Zn down 0 A/cm$^2$ | Zn down 540 A/cm$^2$ |
|---|---|---|---|---|
| Average Zn concentration in the capillaries (at. %) | 3.0 | 4.6 | 8.8 | 13.2 |

Table 2. Ratios of the average concentrations of dissolved Zn measured in the different configurations.

|  | Ratio of the average Zn concentrations |
|---|---|
| **Contribution of convection under electric current** <br><br> Without solutal convection: (Zn up – 782 A/cm$^2$) / (Zn up – 0 A/cm$^2$) <br><br> With solutal convection: (Zn down – 540 A/cm$^2$) / (Zn down – 0 A/cm$^2$) |  <br><br> 1.53 <br><br> 1.50 |
| **Contribution of solutal convection** <br><br> Without current: (Zn down – 0 A/cm$^2$) / (Zn up – 0 A/cm$^2$) <br><br> With current: (Zn down – 540 A/cm$^2$) / (Zn up – 782 A/cm$^2$) |  <br><br> 2.93 <br><br> 2.87 |

In summary, the experimental results show that the contributions of the current and of solutal convection on mass transport are of the same orders of magnitude. This can be analyzed by evaluating quantitatively the solutal, electrostatic and magnetic forces given by the Equ. [2-4] (see quantitative analysis of these forces in the Supplementary materials). For the solutal convection, a force of 590 N/m$^3$ is found, whereas for the electrostatic and magnetic forces under 1000 A/cm$^2$ current density, values of 1.1×10$^{10}$ N/m$^3$ and 460 N/m$^3$ are obtained, respectively. Consequently, the magnetic and solutal forces are in the same range, and below the electrostatic force by orders of magnitude.

Considering these evaluations, it could be concluded that the dominating driving force for mass transport in liquid metals under ≈ 10$^2$-10$^3$ A/cm$^2$ current densities would indisputably be the electrostatic force, which is above the solutal and magnetic forces by about 8 orders of magnitude. However, the electrostatic force being parallel to the electric field and therefore to the axis of the capillary, the liquid metal exerts a force on the closed ends of the capillary. According to Newton's third law, the ends of the capillaries exert then a reaction force opposing the electrostatic force, so that the resulting force is zero. Therefore, the net force (electrostatic force + reaction force) exerted on the liquid as a whole is null. On the contrary, as also described in the Supplementary materials section, the magnetic forces induce poloidal motion of the liquid around the toroidal lines of the magnetic field [9, 10]. In this case, the liquid is free to move along closed poloidal paths, without opposition by reaction forces. Therefore, motion of the liquid can be activated under much lower driving forces, of the order of the solutal force (about 500 N/m$^3$), which is also the typical order of magnitude of the magnetic force. Note that, under low current densities (≈ 0.1 A/cm$^2$), this is the magnetic force which is considered, rather than the electrostatic force [8-10, 16]. The present work shows that this is also the case under current densities of ≈ 10$^2$-10$^3$ A/cm$^2$.

Therefore, this study shows that, in liquid metals under ≈ 10$^2$-10$^3$ A/cm$^2$ current densities in thin capillaries (≈ 1 mm in diameter) limiting convection, mass transport is controlled by the magnetic force. As a result, in most situations of electrotransport in liquid metals (from 0.1 A/cm$^2$ [8, 10, 16] to 10$^2$-10$^3$ A/cm$^2$ [3-5, 17] current densities, convection being controlled by capillaries [3-5, 17], or not controlled as in liquid metal batteries [8-10]), it is the magnetic contribution through Lorentz forces that must be taken into account, rather than the electrostatic contribution typically considered in classical [3-5, 17] and recent [6] works.

**Acknowledgements**

**References**

[1] M. Gerardin, De l'action de la pile sur les sels de potasse et de soude et sur les alliages soumis à la fusion ignée, Comptes Rendus de l'Académie des Sciences 53 (1861) 727-730.

[2] J. Verhoeven, Electrotransport in metals, Metallurgical Reviews 8 (1963) 311-367.

[3] D. Agnoux, J.M. Augeard, J.M. Escanyé, M. Gerl, Electrodiffusion and electroconvection in SnIn and SnSb dilute liquid alloys, Physical Review B 18 (1978) 537-544.

[4] A. Lodding, Electrotransport and effective self-diffusion in pure liquid gallium metal, J. Phys. Chem. Solids 28 (1967) 557-568.

[5] J.L. Blough, D.L. Olson, D.A. Rigney, Electrotransport in Liquid Na-K Alloys, Materials Science and Engineering 11 (1973) 73-79.

[6] D.K. Belashchenko, The relationship between electrical conductivity and electromigration in liquid metals, Dynamics 3 (2023) 405-424.

[7] R.E. Shaw and J.D. Verhoeven, Convection effects during electrotransport of liquid metals, Metallurgical Transactions 4 (1973) 2349-2355.

[8] D.H. Kelley, D.R. Sadoway, Mixing in a liquid metal electrode, Physics of Fluids 26 (2014) 057102.

[9] D.H. Kelley, T. Weier, Fluid mechanics of liquid metal batteries, Applied Mechanics Reviews 70 (2018) 020801.

[10] W. Herreman, S. Bénard, C. Nore, P. Personnettaz, L. Cappanera, J.L. Guermond, Solutal buoyancy and electrovortex flow in liquid metal batteries, Physical Review Fluids 5 (2020) 074501.

[11]. Y. Adda, J. Philibert, La diffusion dans les solides, Institut national des sciences et techniques nucléaires, Saclay, 1966.

[12] P. Personnettaz, S. Landgraf, M. Nimtz, N. Weber, T. Weier. Mass transport induced asymmetry in charge/discharge behavior of liquid metal batteries. Electrochemistry Communications 105 (2019) 106496.

[13]. R.B. Bird, W.E. Stewart, E.N. Lightfoot, Transport phenomena, John Wiley and Sons, New-York, 2002.

[14] Z. Trzaska, J.P. Monchoux, Electromigration experiments by spark plasma sintering in the silver-zinc system, Journal of Alloys and Compounds 635 (2015) 142-149.

[15]. C.H. Ma, R.A. Swalin, A study of solute diffusion in liquid tin, Acta Metallurgica 8 (1960) 388-395.

[16] K. Wang, K. Jiang, B. Chung, T. Ouchi, P.J. Burke, D.A. Boysen, D.J. Bradwell, H. Kim, U. Muecke, D.R. Sadoway, Lithium-antimony-lead liquid metal battery for grid-level storage, Nature 514 (2014) 348-350.

[17] D. Stroud, Calculations of the average driving force for electromigration in liquid-metal alloys, Physical Review B 13 (1976) 4221-4226.

[18] J. Pstrus, Z. Moser, W. Gasior, A. Debski, Surface tension and density measurements of liquid Sn-Zn alloys. Experiment vs. Surdat database of Pb-free solders, Archives of Metallurgy and Materials 51 (2006) 335-343.

[19] T. R. Anthony, The electromigration of liquid metal inclusions in Si, Journal of Applied Physics 51 (1980) 6356-6365.

[20] C.L.Chien, C.R .Westgate, The Hall effect and its applications, Springer Science + Business Media, New-York, 1980.

[21] L. Sani, C. Petrillo, F. Sacchetti, Determination of the interstitial electron density in liquid metals: basic quantity to calculate the ion collective-mode velocity and related properties, Physical Review B 90 (2014) 024207.

[22] J.O. Gasser, J.L. Bretonnet, A. Bruson, Temperature dependence of the liquid tin resistivity, Phys. Stat. Sol. 128 (1985) 789-796.

[23] R.A. Matula, Electrical resistivity of copper, gold, palladium and silver, J. Phys. Chem. Ref. Data 8 (1979) 1147-1298.

[24] A.R. Grone, Current-induced marker motion in copper, J. Phys. Chem. Solids 20 (1961) 88-93.

[25] D. Jiles, Introduction to magnetism and magnetic materials, CRC Press, Boca Raton, 2016.

**Supplementary materials**

- Characterization of the concentration profiles
- Expressions for solutal, electromigration and Lorentz forces (citing Refs. [3, 9-11, 13, 17-25]).